\documentclass[11pt]{article}

\usepackage{xcolor}

\usepackage{amsmath}
\usepackage{amssymb}
\usepackage{graphicx,scalerel}
\usepackage{extsizes}
\usepackage{array}
\usepackage{url}
\usepackage{natbib}
\usepackage{bm}
\usepackage{caption}
\usepackage{float}
\usepackage{enumitem} 
\usepackage{subcaption}
\usepackage{booktabs}
\usepackage{caption}
\usepackage{lipsum}
\usepackage{framed}   
\usepackage{multirow} 
\usepackage{makecell}
\newcommand{\bmath}[1]{\bm{#1}}

\def\bSig\mathbf{\Sigma}

\newcommand{\T}{\mathrm{T}}
\newcommand{\bx}{\bmath{x}}
\newcommand{\bbeta}{\bmath{\beta}}
\newcommand{\btheta}{\bmath{\theta}}
\newcommand{\bgamma}{\bmath{\gamma}}
\newcommand{\bGamma}{\bmath{\Gamma}}
\newcommand{\bff}{\bmath{f}}
\newcommand{\bbb}{\bmath{b}}
\newcommand{\bc}{\bmath{c}}
\newcommand{\bz}{\bmath{z}}
\newcommand{\by}{\bmath{y}}
\newcommand{\bepsilon}{\bmath{\epsilon}}
\newcommand{\bmu}{\bmath{\mu}}

\newcommand{\bzero}{\mathbf{0}}
\newcommand{\be}{\mathbf{e}}
\newcommand{\bj}{\mathbf{j}}
\newcommand{\bd}{\mathbf{d}}
\newcommand{\rank}{\mathrm{rank}}

\AtBeginDocument{}

\usepackage{amsthm}
\newtheorem{lemma}{Lemma}

\usepackage{fullpage}

\begin{document}

\begin{center}
{\large{\textbf{Optimal design of dynamic experiments for scalar-on-function linear models \\
with application to a biopharmaceutical study}} } \\[1ex] 
		
D. Michaelides\footnote{Damianos Michaelides; damianosm@cing.ac.cy; Biostatistics Unit, The Cyprus Institute of Neurology and Genetics, 1683 Nicosia, Cyprus.}
, M. Adamou, D. C. Woods \& A. M. Overstall \\[1ex]

Biostatistics Unit, The Cyprus Institute of Neurology and Genetics, Cyprus \\
Southampton Statistical Sciences Research Institute, University of Southampton, UK
\end{center}			
				

\normalsize

A Bayesian optimal experimental design framework is developed for experiments where settings of one or more variables, referred to as profile variables, can be functions. For this type of experiment, a design consists of combinations of functions for each run of the experiment. Within a scalar-on-function linear model, profile variables are represented through basis expansions. This allows finite-dimensional representation of the profile variables and optimal designs to be found. The approach enables control over the complexity of the profile variables and model. The method is illustrated on a real application involving dynamic feeding strategies in an Ambr250 modular bioreactor system.


\begin{flushleft}
Basis functions; Design of experiments; Functional linear model; Profile variables; Dynamic experiments.
\end{flushleft}


\section{Introduction} \label{s:intro}

In a variety of experiments, especially in biological studies, there can be more flexibility over the control of design variables than is the norm in standard experimental designs. In particular, it is becoming increasingly common to run studies where the value taken by one or more \textit{profile variables} can be adjusted during a single run of the experiment. Hence, for these variables optimal design concerns how to vary their values through the choice of a function for each variable in each run of the experiment. We refer to such examples as \textit{dynamic experiments}.  

Previous research on the design and analysis of dynamic experiments can mostly be classified under one of two different frameworks.

\begin{itemize}
\item[(i)] Optimal design for dynamic mechanistic models, usually derived as the solution to a set of differential equations, which naturally incorporate time-evolving behavior. Early work was reviewed by \citet{titterington1980}. Further developments have often come from the field of control theory (e.g., \citealp{em1989}), especially applications in chemical and biological processes where dynamical systems abound (e.g., \citealp{bi2018}). The methods and designs are often quite closely tailored to the specific experiment or model being studied. 
\item[(ii)] Response surface methodology, typically using standard designs. \citet{georgakis2013} proposed the extension of standard response surface methodology (RSM) to include profile variables. Designs were obtained through application of standard RSM or optimal designs to the so-called sub-factors defined by a dimension reduction of the profile variables (see Section~\ref{s:flmdevelopment}). \citet{roche2018} applied related methods to nuclear safety experiments and extended the approach to include data-driven dimension reduction, e.g., principle components analysis. These ideas have now seen fairly wide-spread adoption in a variety of pharmaceutical applications, see \citet{georgakis2024}, including extensions to experiments with functional responses. 
\end{itemize}


Our work fits within this second framework, and is motivated by biopharmaceutical process development via experiments using an Ambr250 modular bioreactor system. Such a system allows process conditions to be individually controlled for each reaction. A typical experiment would measure the concentration of the product of interest from each reaction (run of the experiment) after a fixed time period, usually of the order of two weeks.    

Design of experiments methods have seen considerable uptake on such systems; for example, \citet{tai2015} applied a definitive screening design to perform variable screening. However, previous applications of statistically designed experiments in such systems have only employed traditional \textit{static} variables, whose values are set and kept constant through each individual run. Several studies have, though, highlighted the potential benefits of dynamic experimentation. \citet{yoon2003}, \citet{trummer2006} and \citet{rameez2014} all found that varying the temperature during a reaction can increase concentration. \citet{lu2013} discussed the increase in concentration that could be obtained from a dynamic feeding strategy in place of the usual fixed bolus feeding. However, none of these studies used experiments specifically designed to estimate the effects of profile variables.

To study the applicability of such suggestions to their own studies, our collaborators at GlaxoSmithKline (GSK) wanted to design and run a series of dynamic experiments. We use one of these experiments to demonstrate our novel approach to designing such studies. The practicalities of the experiment involved varying three static variables: initial viable cell concentration, pH, and temperature, and one profile variable: feed volume. The study aimed to investigate the titre content of the product (concentration), with the eventual aim of optimizing cell growth.

In this paper, we present a novel Bayesian approach to designing experiments with profile variables, assuming a scalar-on-function linear model (Section~\ref{s:flmdevelopment}). Optimal designs are found exploiting the connections between this model and standard linear model theory (Section~\ref{sec:doemethodology}). The impact of the model on the resulting designs is explored via illustrative examples (Section~\ref{s:example}) and methods demonstrated via application to the Ambr250 experiment (Section~\ref{s:application}).

\section{Scalar-on-function linear models and basis representations} \label{s:flmdevelopment}

\subsection{Scalar-on-function linear models}

Suppose there are $J$ profile variables that can be controlled over a time interval $\mathcal{T}=[0,T]$, where $x_j(t)$ denotes the value of the $j$th profile factor at time $t \in \mathcal{T}$. For $j = 1, \dots,J$, it is assumed that $x_j(\cdot) \in \mathcal{X}_j \subset \mathcal{L}^2(\mathcal{T})$, where $\mathcal{L}^2(\mathcal{T})$ is the set of all square integrable functions on $\mathcal{T}$. The sets $\mathcal{X}_1,\dots,\mathcal{X}_J$ are determined by the nature of the experiment and the profile variables. For example, if the $j$th profile variable cannot be changed over $\mathcal{T}$, then $\mathcal{X}_j$ will be the set of constant functions. Therefore, a standard, static experiment with fixed controllable variables is a special case.

The experiment consists of $n$ runs. For $i=1,\dots,n$, the $i$th run consists of specifying the $J$ controllable profile variables, $\bx_i(t) = \left[ x_{i1}(t), \dots, x_{iJ}(t) \right]^{\T} \in \mathcal{X} = \mathcal{X}_1 \times \dots \times \mathcal{X}_J$, and measuring the scalar response, $y_i$, at time $t=T$. 

To investigate the effect of the $J$ profile variables on the response, we assume a scalar-on-function linear model \citep[e.g.,][pages 261-277]{ramsay2005}. Specifically, it is assumed
\begin{equation}
y_i = \int_{\mathcal{T}} \bff \left[ \bx_i(t) \right]^\T \bbeta(t) \; \mathrm{d}t + \epsilon_i\,,
\label{eq:polynomialmodelfunctional}
\end{equation}
for $i=1,\dots,n$. In (\ref{eq:polynomialmodelfunctional}), $\bff[\bx(t)] = \left\{f_1[\bx(t)],\dots,f_Q[\bx(t)]\right\}^\T$ is a vector of regression functions of the profile variables, controlling the complexity of the model, where it is assumed that $f_q : \mathcal{X} \to \mathcal{G}_q \subset \mathcal{L}^2(\mathcal{T})$. For example, a first-order model with an intercept has $\bff[\bx(t)] = [1,\bx(t)]^\T$. Furthermore, $\bbeta(t) = \left[\beta_1(t),\dots, \beta_Q(t) \right]^\T$ is a $Q \times 1$ vector of unknown functional parameters. It is assumed that $\beta_q(t) \in \mathcal{B}_q \subset \mathcal{L}^2(\mathcal{T})$. Lastly, $\epsilon_1,\dots,\epsilon_n$ are random errors, which are assumed independent, with expectation $\mathrm{E}(\epsilon_i) = 0$ and variance $\mathrm{var}(\epsilon_i) = \sigma^2 < \infty$, for $i=1,\dots,n$. The interpretation of the functional parameter $\beta_q(t)$, for $q=1,\dots,Q$, are that the times with large absolute value of $\beta_q(t)$ have highest influence on the response \citep{reiss2017}.

The model given by (\ref{eq:polynomialmodelfunctional}) with univariate functional parameters will often provide a suficiently flexible representation of the data-generating process. It is a special case of the multivariate functional parameter model of \cite{yao2010}; see Appendix A. However, multivariate functional parameters are not straightforward to interpret, hence we focus on the simplification to univariate functional parameters.

\subsection{Basis expansions} \label{sec:bases_fun}

\subsubsection{Functional parameter basis expansion} \label{sec:funpara}

The functional parameters $\beta_1(t),\dots, \beta_Q(t)$ in model (\ref{eq:polynomialmodelfunctional}) are infinite-dimensional. Estimation from a finite number, $n$, of scalar responses, $y_1,\dots,y_n$, can be achieved by assuming a parametric form via a finite basis expansion \citep[][page 44]{ramsay2005}.

For $q=1,\dots,Q$, the basis expansion of $\beta_q(t)$ is
$$
\beta_q(t) = \sum_{l=1}^{m_{\beta,q}} \theta_{ql} b_{ql}(t) = \bbb_q(t)^\T \btheta_q,
$$
where the functions $b_{q1}(t),\dots,b_{qm_{\beta,q}}(t)$ are known basis functions and $\btheta_q$ is an $m_{\beta,q} \times 1$ vector of unknown coefficients. Consequently, the problem is reduced to estimating $p = \sum_{q=1}^Q m_{\beta,q}$ unknown coefficients given by $\btheta = (\btheta_1,\dots,\btheta_Q)^\T$. A special case of a scalar parameter, $\theta_q$, is represented by the single ($m_{\beta,q}=1$) basis function $b_q(t) = 1$. We can write $\bbeta(t) = B(t)^\T \btheta$, where $B(t)$ is a $p \times Q$ block diagonal matrix with $q$th block given by $\bbb_q(t)$, i.e.
$$B(t) = \left[ \begin{array}{llll}
\bbb_1(t) & 0 & \dots & \dots \\
0 & \bbb_2(t) & 0 & \dots \\
\vdots & 0 & \ddots & \vdots \\
\vdots & \vdots & \dots & \bbb_Q(t) \end{array} \right].
$$

\subsubsection{Profile variable basis expansion} \label{sec:profvar}

In the design of experiments setting, we have the freedom to specify the profile variables $\bx_1(t),\dots,\bx_n(t)$ to optimise gain in information in estimating the functional parameters $\bbeta(t)$, or equivalently coefficients $\btheta$, from the scalar responses $y_1,\dots,y_n$. Similar to Section~\ref{sec:funpara}, $\bx_i(t)$, for $i=1,\dots,n$, are infinite-dimensional. To make the optimisation problem tractable, we again use a finite basis expansion. For $i=1,\dots,n$ and $j=1,\dots,J$,
\begin{equation}\label{eq:profile_expansion}
x_{ij}(t) = \sum_{l=1}^{m_{x,j}} \gamma_{ijl} c_{j l}(t) = \bgamma_{ij}^\T \bc_j(t)\,,
\end{equation}
where $c_{j1}(t),\dots,c_{jm_{x,j}}(t)$ are known basis functions and $\bgamma_{ij} = (\gamma_{ij1},\dots,\gamma_{ijm_{x,j}})$ are coefficients. Let $\bGamma = (\bgamma_{11},\dots,\bgamma_{nJ})^\T$ be the $n\sum_{j=1}^J m_{x,j} \times 1$ vector of coefficients. 

For observational data, $\bGamma$ are estimated and treated as fixed in the estimation of $\btheta$. Conversely, in design of experiments, $\bGamma$ are specified to optimise gain in information in estimating $\btheta$ (see Section~\ref{sec:doemethodology}). A sufficiently complex basis expansion should be chosen so that an optimal choice of $\bGamma$ leads to the profile variables reconstructed via~\eqref{eq:profile_expansion} providing a good approximation to the optimal infinite-dimensional profile variables.

\subsection{Standard linear model representation of the scalar-on-function linear model} \label{sec:rewrite}

The key to design of experiments for scalar-on-function linear models is the observation that the model given by (\ref{eq:polynomialmodelfunctional}), under the basis expansions of Section~\ref{sec:bases_fun}, can be written in the form of a standard linear model. This observation has been made previously \citep[e.g.][]{reiss2017} for the case where $J=1$ and the regression functional is of the form $\bmath{f}\left[ x_i(t) \right] = \left[1, x_i(t)\right]^\T$. Below we extend this representation to $J>1$, and more general forms of the regression functional. 

The $q$th element, $f_q[\bx_i(t)]$, of the regression functional can be written as $f_q \left[ \bmath{x}_i(t) \right] = \prod_{a \in \mathcal{F}_q} x_{ia}(t)$, where $\mathcal{F}_q$ is a set defining the $q$th term. The members of $\mathcal{F}_q$ belong to the set $\left\{1,\dots,J\right\}$ and can be repeated. For example, the intercept has $\mathcal{F}_q = \varnothing$, the main effect of the $j$th profile factor has $\mathcal{F}_q = \left\{ j \right\}$, the quadratic effect of the $j$th profile factor has $\mathcal{F}_q = \left\{ j, j \right\}$ and the two-way interaction between the $j_1$th and $j_2$th profile factors has $\mathcal{F}_q = \left\{ j_1, j_2 \right\}$.

The model given by (\ref{eq:polynomialmodelfunctional}) can now be written as
\begin{equation}
 y_{i}= \int_\mathcal{T} \sum_{q=1}^Q \left[ \prod_{a \in \mathcal{F}_q} \bmath{\gamma}_{ia}^\mathrm{T} \bmath{c}_a(t) \right]\bmath{b}_q(t)^\mathrm{T} \bmath{\theta}_q \;\mathrm{d}t  + \epsilon_{i}.  
\label{eqn:modelbig}
\end{equation}
Using properties of Kronecker products (denoted by $\otimes$)
\begin{equation}
\prod_{a \in \mathcal{F}_q} \bmath{\gamma}_{ia}^\mathrm{T} \bmath{c}_a(t)
=\left[ \bigotimes_{a \in \mathcal{F}_q} \bmath{\gamma}_{ia}\right]^\mathrm{T} \left[ \bigotimes_{a \in \mathcal{F}_q} \bmath{c}_{a}(t) \right],
\label{eqn:kronprod}
\end{equation}
where $\bigotimes_{a \in \mathcal{F}_q} \bmath{\gamma}_{ia}$ and $\bigotimes_{a \in \mathcal{F}_q} \bmath{c}_{a}(t)$ are both $\prod_{a \in \mathcal{F}_q} m_{x,a} \times 1$ vectors. Note, we use the convention that a Kronecker product over the empty set, $\varnothing$, is one.

Substituting (\ref{eqn:kronprod}) into (\ref{eqn:modelbig}) gives $y_i = \sum_{q=1}^Q \bmath{z}_{iq}^\mathrm{T} \bmath{\theta}_q + \epsilon_i$, where
\begin{equation}
\bz_{iq} = \int_\mathcal{T} \bmath{b}_q(t) \left[\bigotimes_{a \in \mathcal{F}_q} \bmath{c}_{a}(t)^\mathrm{T}\right] \; \mathrm{d}t \left[ \bigotimes_{a \in \mathcal{F}_q} \bmath{\gamma}_{ia} \right],
\label{eqn:ziq}
\end{equation}
is an $m_{\beta,q} \times 1$ vector. Therefore, the model can be written in the form of a standard linear model as
$\by = Z \btheta + \bepsilon$, where $\by=(y_1,\dots,y_n)$, $\bepsilon = (\epsilon_1,\dots,\epsilon_n)$, and $Z$ is an $n \times p$ model matrix given by
$$Z = \left( 
\begin{array}{lcl}
\bz_{11}^\mathrm{T} & \dots & \bz_{1Q}^\mathrm{T} \\
\vdots & \ddots & \vdots \\
\bz_{n1}^\mathrm{T} & \dots & \bz_{nQ}^\mathrm{T} 
\end{array}\right).$$
Note that $Z$ is a function of the coefficients, $\bGamma = (\bgamma_{11},\dots,\bgamma_{nJ})^\T$, of the basis expansion of the profile variables. However, this dependence is suppressed in the notation for clarity.

From (\ref{eqn:ziq}), $\bz_{iq} = R_q \bigotimes_{a \in F_q} \bmath{\gamma}_{ia}$ where $R_q = \int_{\mathcal{T}} \bbb_q(t) \bigotimes_{a \in \mathcal{F}_q}\bc_a(t)^\T \; \mathrm{d}t$ is an $m_{\beta,q} \times \prod_{a \in \mathcal{F}_q} m_{x,a}$ matrix. The elements of $R_q$ are integrals of products of basis functions. The exact form of the integrands will depend on the choice of basis functions for both the $\bx_i(t)$'s and $\bbeta_q(t)$'s. However, for any combination of monomial and B-spline basis functions, we show, in Appendix B, that the integration can be performed in closed form. We also provide a closed form expression for the elements of $R_q$.

\section{Optimal design for scalar-on-function linear models} \label{sec:doemethodology}

\subsection{Basis expansions and constraints} \label{sec:constraints}

The aim is to specify the vector of profile variable coefficients $\bGamma$ (subsequently referred to as the \emph{design}) to optimise gain in information in estimating the unknown functional parameter coefficients $\btheta$. This will be achieved by minimising an objective function $\Psi(\bGamma)$. The choice of objective function will be addressed in Section~\ref{sec:objective}.
 
However, beforehand, there are several issues that need discussion, in regards to the basis expansion of both the functional parameters and profile variables, and the complexity of the model specified by the regression functional, $\bff(\cdot)$. These matters will need addressing before the experiment, and therefore can be considered as part of the design process.

The basis functions, $\bbb_1(t),\dots,\bbb_Q(t)$ of the functional parameter expansion need specification. For these functions, we consider monomials and B-splines. Any combination of these, combined with B-splines for the profile variable basis expansion, leads to the integrals in the model matrix $Z$ being available in closed form (as discussed in Section~\ref{sec:rewrite}). Monomial basis functions have the advantage of higher interpretability but B-splines are more computationally efficient. 

The complexity of the scalar-on-function linear model needs to be specified via the regression functional, $\bff(\cdot) = \left[ f_1(\cdot),\dots,f_Q(\cdot)\right]^\T$, and the number of basis functions, $m_{\beta,1},\dots,m_{\beta,Q}$. Specifying the complexity of a statistical model before observing the responses is a challenge shared by all applications of optimal design of experiments \cite[e.g.][page 329]{atkinson2007}. It is a subjective decision which should be made in consultation with the researchers, considering the number of runs, $n$, in the experiment and, potentially, the estimability of $\btheta$. On this last point, for $Z^\T Z$ to be non-singular, and the least squares estimates of $\btheta$ to exist, the $n \times p$ model matrix $Z$ requires full column-rank. A necessary, but not sufficient, condition for this to occur is that $p = \sum_{q=1}^Q m_{\beta,q} \le n$. This places an upper bound on the complexity ($m_{\beta,1},\dots,m_{\beta,Q}$) of the basis expansion of the functional parameters, based on the available experimental resources ($n$). 

Similarly, the functions, $\bc_1(t),\dots,\bc_J(t)$, and numbers, $m_{x,1},\dots,m_{x,J}$, of the profile variable basis expansion need specification. Ultimately, there will be a compromise between specifying profile variables $\bx_1(t),\dots,\bx_n(t)$ that are simple enough to be implemented in a physical experiment but complex enough to provide a good approximation to the actual optimal infinite-dimensional profile variables. The former can be determined in consultation with the researchers. The latter can be investigated as part of minimising the objective function $\Psi(\cdot)$ (see Section~\ref{s:example} and \ref{s:application} for examples).

For the basis functions, $\bc_1(t),\dots,\bc_J(t)$, the proposal is to use B-splines. These offer two distinct advantages. Firstly, in combination with B-spline or monomial basis functions for the functional parameter basis expansions, the integrals in the model matrix are available in closed form. Secondly, as shown below, it is straightforward to implement constraints on the profile variables $\bx_i(t)$, for $i=1,\dots,n$, via simple constraints on the elements of design $\bGamma$.

Often, the profile variable $x_{ij}(t)$, for $i=1,\dots,n$ and $j=1,\dots,J$, will be constrained in the experiment such that $x_{ij}:\mathcal{T} \to [a_j,b_j]$. For example, if $x_{ij}(t)$ represented the temperature at time $t \in \mathcal{T}$ for the $i$th run, then $a_j$ and $b_j$ are the minimum and maximum possible temperatures, respectively.

Suppose, for $j=1,\dots,J$, the basis functions $c_{j1}(t),\dots,c_{jm_{x,j}}(t)$ are B-spline functions. Using the property of B-spline functions that $\sum_{l=1}^{m_{x,j}} c_{jl}(t) = 1$ for $t \in \mathcal{T}$, if the elements of $\bgamma_{ij}$ are such that $\gamma_{ijl} \in [a_j,b_j]$, for all $l=1,\dots,m_{x,j}$, it follows that $x_{ij}(t) = \sum_{l=1}^{m_{x,j}} \gamma_{ijl} c_{jl}(t) \in [a_j,b_j]$ for $t \in \mathcal{T}$.

To implement B-spline basis functions for $x_{ij}(t)$, the number, $k_{j}$, and location of the internal knots need to be specified. It is common to choose uniformly spaced internal knots and we use this approach in this paper. However, the location of the knots could be incorporated into the design problem. For maximally smooth basis functions, $m_{x,j} = k_j + d_j + 1$, where $d_j$ is the degree. Thus the number of internal knots is specified by the number of basis functions, $m_{x,j}$, and the degree.


The following lemma can aid in choosing the number of basis functions. It gives a lower bound on the complexity of the profile variables for the chosen complexity of the scalar-on-function linear model. The proof is provided in Appendix C.  

\begin{lemma} \label{lemma:1}
A necessary, but not sufficient condition, for the least squares estimates of the functional parameter coefficients, $\btheta$, to exist, is $m_{\beta,q} \le \prod_{a \in \mathcal{F}_q} m_{x,a}$, for all $q=1,\dots,Q$.
\end{lemma}

\color{black}

\subsection{Objective functions via Bayesian decision-theory} \label{sec:objective}

We formulate objective functions to find designs by using a Bayesian decision-theoretic approach. This approach has the advantages that the experimental goal can be incorporated via the choice of loss function and all assumed sources of uncertainty are accounted. The post-experiment analysis does not necessarily need to be undertaken under a Bayesian approach.

Bayesian decision-theoretic design (or Bayesian design) starts with specification of a loss function, denoted $\ell(\btheta,\by,\bGamma)$, giving the loss in estimating parameters $\btheta$, using responses $\by$ obtained from an experiment with design $\bGamma$. The objective function, $\Psi(\bGamma)$, follows from the expected loss $L(\bGamma) = \mathrm{E}_{\by, \btheta \vert \bGamma} \left[\ell(\btheta,\by,\bGamma)\right]$, where expectation is with respect to the joint probability distribution of the responses $\by$ and parameters $\btheta$. This joint probability distribution of $\by$ and $\btheta$ is given by the distribution of the responses $\by$ given $\btheta$, and the prior distribution of $\btheta$. Note that the expected loss can always be written as $L(\bGamma) = g \left[ \Psi(\bGamma) \right]$, where $g(\cdot)$ is a monotonically increasing function. Therefore, it is sufficient to minimise the objective function $\Psi(\bGamma)$ in place of $L(\bGamma)$.

The Bayesian design is given by minimising the objective function $\Psi(\bGamma)$ over the space of all designs. Recall from Section~\ref{sec:constraints}, that if the profile variables are constrained such that $x_{ij}:\mathcal{T} \to [a_j,b_j]$, for  $i=1,\dots,n$ and $j=1,\dots,J$, then it is sufficient for $\bgamma_{ijl} \in [a_j,b_j]$, for $l=1,\dots,m_{x,j}$, $i=1,\dots,n$ and $j=1,\dots,J$. This defines the space of all designs.

For the scalar-on-function linear model, the distribution of $\by$ given $\btheta$ follows from specification of a distribution for the errors $\epsilon_1,\dots,\epsilon_n$. For analytical tractability, we assume that the errors are normally distributed. For $\btheta$, we assume a multivariate normal distribution with mean $\bmu$ and variance $\sigma^2 \Sigma$, where $\bmu$ and $\Sigma$ are a specified $p \times 1$ vector and $p \times p$ matrix, respectively. We also assume an inverse-gamma prior distribution for the nuisance parameter of the error variance $\sigma^2$. The shape and rate parameters are $a/2$ and $b/2$, respectively, where $a$ and $b$ are specified scalar constants.  

Let $\hat{\bmu} = \hat{\Sigma} \left(Z^\T \by + \Sigma^{-1} \bmu \right)$, $\hat{\Sigma} = \left(Z^\T Z + \Sigma^{-1} \right)^{-1}$, $\hat{a} = a + n$ and $\hat{b} = b + \bmu^\T \Sigma^{-1} \bmu + \by^\T \by - \hat{\bmu}^\T \hat{\Sigma}^{-1} \hat{\bmu}$. Under the above prior specification, the marginal posterior distribution of $\btheta$ is a multivariate t-distribution with mean $\hat{\bmu}$, variance $\hat{b}\hat{\Sigma}/(\hat{a}-2)$ and $\hat{a}$ degrees of freedom. 

In Sections~\ref{s:example} and \ref{s:application}, we consider the squared error (SE) loss function given by the sum of squared differences of the elements of $\btheta$ and their marginal posterior means, i.e. 
$$\ell_{SE}(\btheta, \by, \bGamma) = \left( \btheta - \hat{\bmu}\right)^\T \left( \btheta - \hat{\bmu}\right).$$
It follows that the SE objective function given by the expected SE loss is $\Psi_{SE}(\bGamma) = \mathrm{tr} \left( \hat{\Sigma} \right)$. Note that finding the Bayesian SE design does not require the prior mean, $\bmu$, nor the prior shape, $a$, or rate, $b$, parameters to be specified. 

If an improper uniform prior distribution is assumed for $\btheta$, whereby the elements of $\Sigma^{-1}$ are zero, then the SE objective function reduces to $\Psi_{SE}(\bGamma) = \mathrm{tr} \left[ \left( Z^\T Z \right)^{-1} \right]$, and the corresponding Bayesian design is equivalent to the classical A-optimal design. 

As presented above, the loss function is in terms of the functional parameter coefficients $\btheta$. However, these may not be of direct interest. Rather, interest may lie in implied functions $\bbeta(t) = \left[ \beta_1(t),\dots,\beta_Q(t)\right]^\T$. Therefore, we propose an alternative loss, in terms of $\btheta$, that is given by integrating a loss, in terms of $\bbeta(t)$, over $\mathcal{T}$. 

As an example, consider the SE loss in terms of $\bbeta(t)$ given by
$$\ell_{SE}(\bbeta(t),\by,\bGamma) = \left[\bbeta(t) - \mathrm{E}_{\bbeta(t)\vert \by, \bGamma} \left(\bbeta(t)\right) \right]^\T \left[\bbeta(t) - \mathrm{E}_{\bbeta(t)\vert \by, \bGamma} \left(\bbeta(t)\right) \right],$$
where $\mathrm{E}_{\bbeta(t)\vert \by, \bGamma} \left[\bbeta(t)\right]$ is the marginal posterior mean of $\bbeta(t)$. Using properties of the multivariate t-distribution, the posterior distribution of $\bbeta(t)$, for $t \in \mathcal{T}$, is multivariate t-distribution with posterior mean $\mathrm{E}_{\bbeta(t) \vert \by, \bGamma} \left[ \bbeta(t) \right] = B(t)^\T \hat{\bmu}$. The implied loss in terms of $\btheta$ is now
\begin{eqnarray*}
\ell_{WSE}(\btheta, \by, \bGamma) & = & \int_{\mathcal{T}} \ell_{SE}(\bbeta(t),\by,\bGamma) \; \mathrm{d}t,\\
& = & \int_{\mathcal{T}} \left[ B(t)^\T \btheta - B(t)^\T \hat{\bmu} \right]^\T \left[ B(t)^\T \btheta - B(t)^\T \hat{\bmu} \right] \; \mathrm{d}t\\
& = & \left(\btheta - \hat{\bmu}\right)^\T B_I \left(\btheta - \hat{\bmu}\right),
\end{eqnarray*}
where $B_I = \int_{\mathcal{T}} B(t)B(t)^\T \; \mathrm{d}t$ is a $p \times p$ matrix. Thus the implied loss in terms of $\btheta$ is a weighted squared error (WSE) loss. The corresponding objective function is $\Psi_{WSE}(\bGamma) =  \mathrm{tr}\left[ B_I \hat{\Sigma} \right]$. Under an improper uniform prior distribution for $\btheta$, the WSE Bayesian design is equivalent to a classical L-optimal design \citep[e.g.][Section 10.5]{atkinson2007}.

\section{Illustrative experiment with a single profile variable} \label{s:example}

In this section, we consider an illustrative example where Bayesian designs are found for an experiment with $J=1$ profile variable. The time interval is $\mathcal{T}=[0,1]$, i.e. $T=1$. The regression functional is $\bff\left[x_{i1}(t) \right] = \left[1, x_{i1}(t) \right]$, so that $Q=2$ and the model is $y_i = \beta_1 + \int_0^1 \beta_2(t) x_{i1}(t) \; \mathrm{d}t + \epsilon_i$, for $i=1,\dots,n$. 

For the basis expansion of $x_{i1}(t)$, we consider degree-zero B-spline basis functions with uniformly spaced internal knots. If there are $m_{x,1}$ basis functions, then this amounts to each $x_{i1}(t)$ being a step function on $\mathcal{T}=[0,1]$, where the discontinuities correspond to the internal knots $\lambda_{1,1},\dots,\lambda_{1,k_1}$, where $k_1 = m_{x,1} - 1$. Explicitly, for $l=1,\dots,m_{x,1}$, 
$$c_{1l}(t) = \left\{
\begin{array}{ll}
1 & \mbox{if $t \in [\lambda_{1,l-1},\lambda_{1,l})$;}\\
0 & \mbox{if otherwise;}
\end{array} \right.$$
where $\lambda_{1,0}=1$ and $\lambda_{1,m_{x,1}}=1$.

Step functions are an important and useful tool in practice. For instance, \citet{rameez2014} considered an Ambr bioreactor application with a temperature step shift from the high to the low level. We will vary the number of basis functions, $m_{x,1}$, which is equivalent to varying the complexity of the profile variable, to investigate the effect on the performance of the design through the expected loss.

For the basis expansion of the functional parameters, $\beta_1$ is a constant so $m_{\beta,1}=1$ and $b_1(t)=1$. For $\beta_2(t)$, we consider two cases with monomial basis functions. In Case I, $m_{\beta,2}=2$, $\bbb_2(t) = (1,t)^\T$, giving $p=3$. In Case II, $m_{\beta,2}=3$, $\bbb_2(t) = (1,t,t^2)^\T$, giving $p=4$. 

For both cases, we find SE and WSE Bayesian designs with number of basis functions for the profile variable $m_{x,1}=2 \mbox{ (not Case II)}, 3, 4, 8,16,100$ and $n=4,12$. Note that $m_{x,1}=2$ is forbidden for Case II since the least squares estimate of $\btheta$ does not exist for this value (see Lemma~\ref{lemma:1}). Section SM4 gives expressions for the two matrices, $R_1$ and $R_2$, which feature in the model matrix $Z$, and the weighting matrix, $B_I$, for WSE Bayesian designs.

The design, $\bGamma$, is a vector with $n\sum_{j=1}^J m_{x,j} = nm_{x,1}$ elements. The profile variable $x_{i1}(t)$ is assumed to be constrained to $[-1,1]$ meaning the elements of $\bGamma$ are also constrained to $[-1,1]$ (see Section~\ref{sec:constraints}). In each scenario, the coordinate exchange algorithm \citep{meyer1995} is used to minimise the objective function from 1000 random starts, as suggested by \citet{goos2011}.

\begin{table}[b]
\centering
\footnotesize
\caption{SE and WSE objective function values for optimal designs for $m_{x,1}=2,3,4,8,16,100$ and $n=4,12$. The profile variables for the scenario highlighted in bold are shown in Figure~\ref{fig:linearstep1}.}
\label{tab:toyexample1}

\begin{subtable}{\linewidth}
\centering
\caption{SE objective function values}
\begin{tabular}{rrrrr} \toprule
$m_{x,1}$ & \multicolumn{2}{c}{Case I} & \multicolumn{2}{c}{Case II} \\
         & $n=4$ & $n=12$ & $n=4$ & $n=12$ \\ \midrule
2   & 8.750  & 2.583 & -       & -     \\
3   & 8.828  & 2.778 & 386.408 & 126.409 \\
4   & 8.750  & 2.570 & 246.869 & 67.735 \\
\textbf{8}   & \textbf{8.493}  & 2.539 & 218.479 & 65.217 \\
16  & 8.427  & 2.520 & 208.843 & 63.610 \\
100 & 8.404  & 2.512 & 206.884 & 63.028 \\ \bottomrule
\end{tabular}
\end{subtable}

\vspace{1ex}

\begin{subtable}{\linewidth}
\centering
\caption{WSE objective function values}
\begin{tabular}{rrrrr} \toprule
$m_{x,1}$ & \multicolumn{2}{c}{Case I} & \multicolumn{2}{c}{Case II} \\
         & $n=4$ & $n=12$ & $n=4$ & $n=12$ \\ \midrule
2   & 1.417 & 0.472 & -     & -     \\
3   & 1.581 & 0.499 & 3.363 & 1.120 \\
4   & 1.417 & 0.472 & 3.243 & 1.022 \\
8   & 1.417 & 0.472 & 3.147 & 1.021 \\
16  & 1.417 & 0.472 & 3.099 & 1.016 \\
100 & 1.417 & 0.472 & 3.094 & 1.010 \\ \bottomrule
\end{tabular}
\end{subtable}
\end{table}

Table~\ref{tab:toyexample1} shows the values of the SE and WSE objective function for the respective optimal designs for each value of $m_{x,1}$ and $n$. Clearly as $n$ increases, the expected loss decreases. Typically, as the number of basis functions, $m_{x,1}$, increases, the expected loss decreases, since the finite-dimensional representation of the profile variable becomes a better approximation to the optimal infinite-dimensional profile variable. The rate of this decrease slows as $m_{x,1}$ increases allowing a compromise to be made between profile variable complexity and gain in information. The exception to expected loss decreasing with $m_{x,1}$ is for $m_{x,1}=2$ for Case I. The uniformly-spaced internal knots for $m_{x,1}=2,3,4$ are $\left\{\frac{1}{2}\right\}$, $\left\{\frac{1}{3},\frac{2}{3}\right\}$ and $\left\{\frac{1}{4},\frac{1}{2},\frac{3}{4}\right\}$, respectively. A change at the knot at $t=\frac{1}{2}$ is important for small $m_{x,1}$ and its absence for $m_{x,1}=3$, leads to higher expected loss than for $m_{x,1}=2$.

\begin{figure}[t]
\begin{center}
\centerline{\includegraphics[width=15cm]{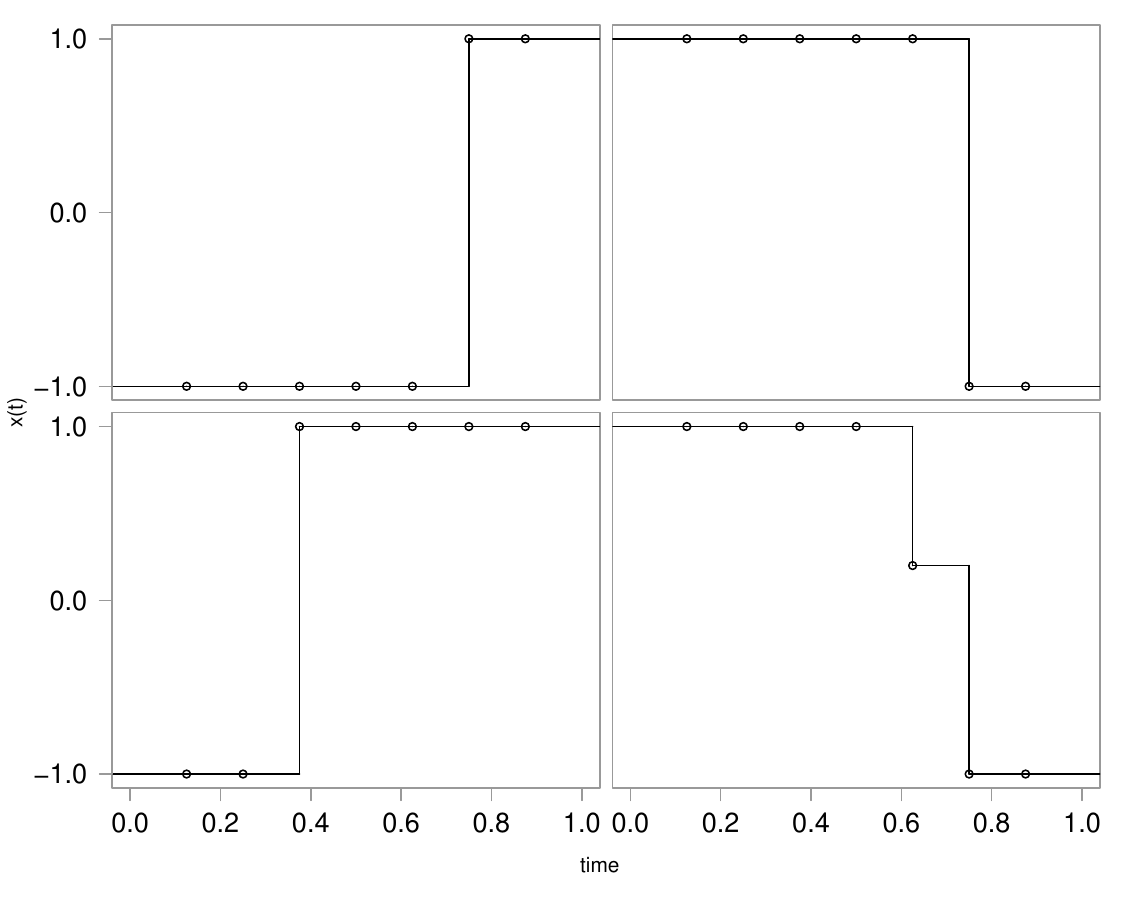}}
\end{center}
    \caption{Profile variables for the $n=4$ SE design for Case I with $m_{x,1}=8$ basis functions.} \label{fig:linearstep1}
\end{figure}

As an example, Figure~\ref{fig:linearstep1} shows the profile variables $x_{11}(t),\dots,x_{41}(t)$ for the $n=$ SE design for Case I with $m_{x,1}=8$ basis functions. This scenario is highlighted in bold in Table~\ref{tab:toyexample1}. Three profile variable have exactly one change and the remaining has exactly two changes.

\color{black}

\section{Application to a dynamic experiment in the Ambr250 bioreactor} \label{s:application}

We now apply the methodology to a real Ambr250 bioreactor experiment. GSK are investigating the relationship between titre content of a product (constrained to $(0,1)$) and $J=4$ variables: feed volume (denoted $x_{i1}(t)$); initial viable cell concentration ($x_{i2}$); pH ($x_{i3}$); and temperature ($x_{i4}$). Feed volume is a profile variable while the remaining variables are static, i.e. $m_{x,2}=m_{x,3}=m_{x,4}=1$. The response, $y_i$, is the logit of titre content. GSK are able to run a dynamic experiment on the (scaled) time interval $\mathcal{T}=[0,1]$ with $n=12$ runs. 

In consultation with GSK, the following regression functional is proposed
$$\bff[\bx_{i}(t)] = \left[1, x_{i1}(t), x_{i2},x_{i3},x_{i4},x_{i2}^2,x_{i3}^2,x_{i4}^2\right]^\T,$$
giving $Q=8$. For all $q=1,\dots,Q$, apart from $q=2$, $m_{\beta,q}=1$ giving constant parameters. For the functional parameter $\beta_2(t)$, we consider the monomial basis function $\bbb_2(t) = (1,t,t^2)^\T$, i.e. $m_{\beta,2}=3$. This leaves a model with $p=\sum_{q=1}^Q m_{\beta,q} =10$.

For the profile variable $x_{i1}(t)$, for $i=1,\dots,n$, GSK are able to implement step functions with at most three changes during $\mathcal{T}$. Thus we consider zero-degree B-splines with $m_{x,1}=4$ (allowing up to three changes). For simplicity, we use uniformly-spaced internal knots.

For the matrices $R_1,\dots,R_Q$ that feature in the model matrix, following the same derivation as in Appendix D, $R_q=1$, for $q \ne 2$, are all scalars, and $R_2$ is a $m_{\beta,2} \times m_{x,1} = 3 \times 4$ matrix given by 
$$R_2 = \frac{1}{192} \left( \begin{array}{cccc}
48 & 48 & 48 & 48 \\
6 & 18 & 30 & 42 \\
1 & 7 & 19 & 37 
\end{array} \right).
$$

It was decided that focus would be on estimating the $\btheta$ parameters rather than $\bbeta(t)$. Therefore we find a SE Bayesian design, which is equivalent to a classical A-optimal design.

\begin{figure}[t]
\begin{center}
\centerline{\includegraphics[width=15cm]{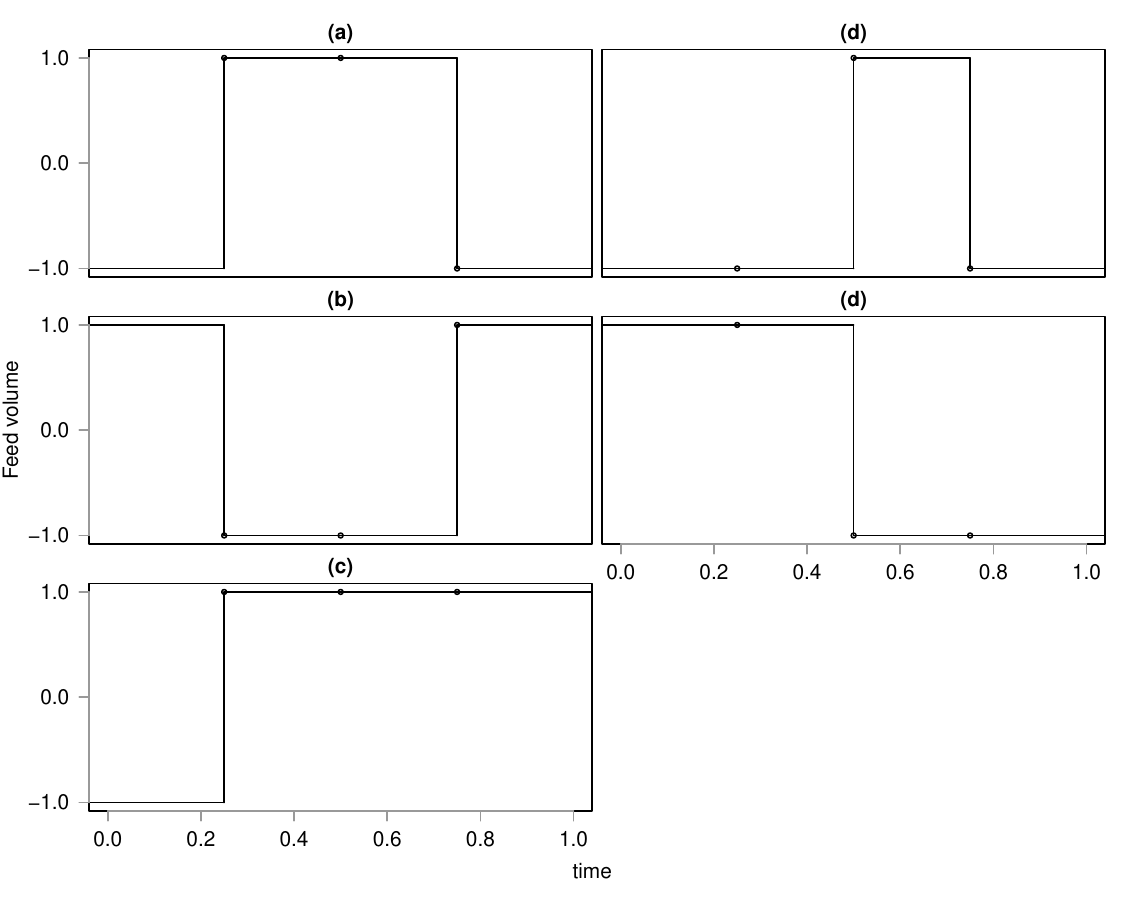}}
\end{center}
    \caption{Five unique Feed volume profile variables of the SE Bayesian design for the Ambr250 bioreactor experiment.} \label{fig:ambr250}
\end{figure}

The design $\bGamma$ has a total of $n\sum_{j=1}^J m_{x,j} = 84$ elements. All controllable variables are scaled to $[-1,1]$ meaning elements of $\bGamma$ are constrained to the same interval. The SE Bayesian design is found using coordinate exchange, as in Section~\ref{s:example}. Table~\ref{table:ambr250} shows the optimal design for the $n=12$ runs. Five unique functions for feed volume, $x_{i1}(t)$, were found, labelled (a)-(e) in Table~\ref{table:ambr250} and shown in Figure~\ref{fig:ambr250}. Function (a) is repeated four times, function (b) repeated five times, and functions (c), (d), and (e) once. The optimal design for the scalar variables includes boundary points and centre points in order to be able to estimate the parameters associated with the quadratic terms. This is similar to the behaviour of the quadratic function for the profile variable, where the functions of the profile variable changes at most twice.

As a sensitivity analysis, we find the SE optimal design for $m_{x,1}=3,4,8,16,100$. The value of the objective function is shown in Table~\ref{tab:sens} in Appendix E. There is limited gain in information in increasing $m_{x,1}$ beyond $m_{x,1}=4$. However, if $m_{x,1}=3$, then there is a large increase in the expected loss. From Figure~\ref{fig:ambr250}, the optimal profile variables make at most two changes over $\mathcal{T}$. Therefore, the increase in expected loss for $m_{x,1}=3$ is due to the position of the internal knots rather than their number. 

On running the experiment, there were two complications. Firstly, run $i=7$ failed to return a titre value, so there are only responses from the remaining eleven runs. Secondly, the Ambr250 experimental apparatus used by GSK provides titre content to three decimal places. For run $i=1$, a titre content of one was returned. This means the true observed titre content was in the interval $[0.9995, 1)$, meaning the true observed response $y_1$ (after logit transform) was in the interval $[7.600, \infty)$ (to three decimal places). To address this complication, we assumed that the true response was censored and estimated $\btheta$ (and $\sigma^2$) via maximum likelihood. This means the likelihood contribution for runs $i>1$ is $\phi \left\{ \left(y_i - \int_{\mathcal{T}} \bff[\bx_{i}(t)])^\T \bbeta(t) \mathrm{d}t\right)/\sigma \right\}$, and for run $i=1$ is $1 - \Phi \left\{ \left(7.600 - \int_{\mathcal{T}} \bff[\bx_{i}(t)])^\T \bbeta(t) \mathrm{d}t\right)/\sigma \right\}$, where $\phi(\cdot)$ and $\Phi(\cdot)$ are the probability density and cumulative distribution functions, respectively, of the standard normal distribution.

To investigate the relationship between the variables and titre content, we used model selection. Specifically, the model that minimised the Akaike information criterion (AIC), found using backward selection, retained main effects for initial viable cell concentration (IVCC), pH and temperature and a quadratic effect for temperature, with regression functional $\bff[\bx_i(t)] = (1, x_{i2},x_{i3},x_{i4},x_{i4}^2)^\T$. The profile variable of feed volume was found not to significantly affect titre content. The final estimated fitted model is
\begin{eqnarray*}   
\widehat{\mbox{logit(titre)}}  & = & 1.199 + 1.324 \; \text{pH} +  1.684 \; \text{IVCC}\\
& & \qquad + 0.506 \; \text{Temperature} -1.851 \; \text{Temperature}^{2},
\end{eqnarray*}
from which, in scaled units, optimal titre content is estimated to occur when pH is $x_2=1$, IVCC is $x_3=1$ and temperature is $x_4 = 0.137$ (to three decimal places).

\begin{table}[b]
\centering
\footnotesize
\caption{SE design for the Ambr250 bioreactor experiment. The feed volume profile variables labelled (a)-(e) are shown in Figure~\ref{fig:ambr250}.}
\label{table:ambr250}
\begin{tabular}{rrrrr} \hline 
Run & Feed volume & IVCC & pH & Temperature \\ 
$i$ & $x_{i1}(t)$ & $x_{i2}$ & $x_{i3}$ & $x_{i4}$ \\ \hline
  1 & (a) & 0 & -1 & 1 \\ 
  2 & (a) & 1 & 1 & 1 \\ 
  3 & (b) & 0 & 1 & -1 \\ 
  4 & (c) & 0 & 0 & 0 \\ 
  5 & (d) & 0 & 0 & 0 \\ 
  6 & (a) & -1 & 1 & 0 \\ 
  7 & (e) & 0 & 0 & 0 \\ 
  8 & (b) & 1 & -1 & 0 \\ 
  9 & (b) & -1 & 0 & 1 \\ 
  10 & (a) & 1 & 0 & -1 \\ 
  11 & (b) & 1 & 1 & 0 \\ 
  12 & (b) & 0 & 0 & 1 \\ 
\hline
\end{tabular}
\end{table}

\section{Discussion}
\label{s:discuss}

In this paper, new methodology for finding optimal designs has been proposed for dynamic experiments where a scalar response
depends on profile variables. The methodology uses basis function expansions of the profile variables and the functional parameters in a scalar-on-function linear model. It is flexible and can be applied assuming a variety of different expansions for both profile variables and parameter functions. 

As discussed in Section~\ref{sec:constraints}, the complexity of the model needs to be specified before the experiment. Part of that specification involves the number of basis functions for the expansion of the functional parameters. An approach is to set the number of basis functions as large as possible and to penalise via a roughness penalty \cite[][Chapter 5]{ramsay2005} implemented through the prior variance matrix, $\Sigma$. Details of this are given in Appendix F.

Ongoing work is to extend the current methodology to non-normally distributed responses and finding designs for scalar-on-function generalised linear models. In this case, a challenge will be the objective function given by the expected loss will not be available in closed form. However, it is anticipated that computational advancements in Bayesian optimal design over the last decade \citep[e.g.][]{rainforth_etal_2024} can be employed. 

In this paper it is assumed that time $t\in [0,\mathcal{T}]$ is the continuous single input to profile variables. The methodology can be naturally extended to scenarios where profile variables have multiple inputs, such as in spatio-temporal studies. 

A limitation of Bayesian optimal design is that it requires assumptions about the data-generating process to be made prior to observing responses. Gibbs optimal design \citep{overstall_etal_2025} is an alternative approach to experimental design which has been developed to be less sensitive to such assumptions. This is a further avenue of future research.

\section*{Acknowledgements}

This work was supported by a PhD studentship (DM) and institutional sponsorship (MA) from the U.K. Engineering and Physical Sciences Research Council and the University of Southampton. The authors are grateful to Gary Finka and Douglas Marsh from GlaxoSmithKline who supported the Ambr250 bioreactor experiment.

\newpage
\bibliography{my_paper}

\clearpage
\appendix

\section*{Appendices}

\section{Scalar-on-function linear models with multivariate functional parameters}

From \cite{yao2010}, the $p$th order model (where $p>2$ is an integer and $J=1$ for simplicity) is
\begin{eqnarray}
y_i &=& \alpha + \int_{\mathcal{T}} x_i(t) A(t) \; \mathrm{d}t + \int_{\mathcal{T}}\int_{\mathcal{T}} x_i(t_1)x_i(t_2) A(t_1,t_2) \; \mathrm{d}t_1 \mathrm{d} t_2 \nonumber \\
& & + \int_{\mathcal{T}}\int_{\mathcal{T}}\dots\int_{\mathcal{T}} x_i(t_1)x_i(t_2)\dots x_i(t_p) A(t_1,t_2,\dots,t_p) \; \mathrm{d}t_1 \mathrm{d} t_2 \dots \mathrm{d} t_p, \nonumber \\
& & \qquad + \epsilon_i \label{eqn:ymmodel}
\end{eqnarray}
where $A(t)$, $A(t_1,t_2)$, $\dots$ , $A(t_1,t_2,\dots,t_p)$ are the multivariate functional parameters. From, for example, \cite{balakrishnan2003}, a property of the Dirac delta function $\delta(\cdot)$ is
\begin{equation}
\int_{\mathcal{T}} g(x,y) \delta(y-x) \mathrm{d}y = g(x,x),
\label{eqn:dirac}
\end{equation}
for any function $g(\cdot,\cdot)$. Letting 
\begin{eqnarray*}
A(t_1,t_2) &=& \delta(t_2-t_1)A^*(t_1,t_2)\\
\vdots &=& \vdots \\
A(t_1,t_2,\dots,t_p) &=& \delta(t_p-t_{p-1}) \dots \delta(t_2-t_1)A^*(t_1,t_2,\dots,t_p)
\end{eqnarray*}
and applying the property in (\ref{eqn:dirac}) to the model (\ref{eqn:ymmodel}) gives
\begin{eqnarray*}
y_i & = & \alpha + \int_{\mathcal{T}} x_i(t) A(t) \; \mathrm{d}t + \int_{\mathcal{T}} x_i(t)^2 A^*(t,t) \; \mathrm{d}t \nonumber \\
& & \qquad + \int_{\mathcal{T}} x_i(t)^p A^*(t,t, \dots, t) \; \mathrm{d}t + \epsilon_i . \label{eqn:simple}
\end{eqnarray*}
Then the model in (\ref{eqn:simple}) is equivalent to the model in equation (1) of the main manuscript with $\bff[x_i(t)] = [1, x_i(t), x_i(t)^2, \dots, x_i(t)^p ]$ and 
$$\bbeta(t) = [\alpha, A(t), A^*(t,t),\dots,A^*(t,t,\dots,t)].$$
This derivation can be extended to cases with $J>1$ in a straightforward manner.

\section{Evaluating the model matrix $Z$}

\subsection{Explicit expression for elements of $R_q$}

Suppose the elements of $\mathcal{F}_q$ can be labelled $\left\{a_1,\dots,a_L\right\}$. Then $\bz_{iq} = R_q \otimes_{\ell=1}^L \bgamma_{ia_\ell}$ where the $m_{\beta,q} \times \prod_{\ell=1}^L m_{x,a_\ell}$ matrix $R_q$ has $uv$th element
\begin{equation}
R_{quv} = \int_0^T b_{qu}(t) \prod_{\ell=1}^L c_{a_\ell,s_{\ell}(v)}(t) \; \mathrm{d}t.
\label{eqn:Rq}
\end{equation}
In (\ref{eqn:Rq}), 
$$s_{\ell}(v) = \left\{ 
\begin{array}{ll}
m_{x,a_{\ell}} & \mbox{if $\omega = 0$;}\\
\left\lceil \frac{\omega}{\prod_{j=\ell+1}^{L+1} m_{x,a_j}}\right\rceil & \mbox{if otherwise;}
\end{array}\right.$$
where $\lceil \cdot \rceil$ is the integer ceiling, $\omega$ is the remainder of $v/\prod_{j=\ell}^L m_{x,a_j}$, i.e. $\omega$ is $v$ modulo $\prod_{j=\ell}^L m_{x,a_j}$, and $m_{x,a_{L+1}}$ is defined as one. 

\subsection{Evaluating integrals in $R_q$}

The expression in (\ref{eqn:Rq}) shows that the elements of $R_q$ are integrals of products of basis functions. Suppose the basis functions chosen for the $x_{ij}(t)$'s and $\beta_q(t)$'s are any combination of monomial and/or B-splines. Specifically, suppose that the product in the integrand of (\ref{eqn:Rq}) features a total of $S$ functions from $M$ collections of B-spline basis functions. The $j$th function in the $m$th collection is denoted $\mathfrak{B}_{j,d_m}^{(m)}(t)$, where $d_m$ is the degree, and $j=1,\dots,v_m$ and $m=1,\dots,M$. The internal knots associated with the $m$th collection are denoted $\lambda_{m,1},\dots,\lambda_{m,k_m}$, where $v_m = k_m + d_m + 1$. Finally, the $w$th associated extended knots is denoted $\lambda^*_{m,w}$, where the extended knots are
$$\left( \underbrace{\lambda_{m,1}, \dots,\lambda_{m,1}}_{d_m+1}, \lambda_{m,2}, \dots, \lambda_{m,k_m-1}, \underbrace{\lambda_{m,k_m}, \dots,\lambda_{m,k_m}}_{d_m+1}\right)^\T.$$

Then, in general, the integral in (\ref{eqn:Rq}) can be written
$$I(r, \bm, \bj, \bd) = \int_{\mathcal{T}} t^r \prod_{s=1}^S \mathfrak{B}_{j_s,d_{m_s}}^{(m_s)}(t) \; \mathrm{d}t,$$
where $r \ge 0$ is an integer, and $\bm = (m_1,\dots,m_S)^\T$, $\bj = (j_1,\dots,j_S)^\T$, and $\bd = (d_{m_1},\dots,d_{m_S})^\T$ identify the collections, the functions within and degree of each collection, respectively, in the product.

The recursive formulation for B-spline basis functions \citep{deboor1978} is
$$\mathfrak{B}_{j_s,d_{m_s}}^{(m_s)}(t) = \frac{t - \lambda^*_{m_s,j_s}}{\alpha(m_s,j_s,d_s)} \mathfrak{B}_{j_s,d_{m_s}-1}^{(m_s)}(t) + 
\frac{\lambda^*_{m_s,j_s+d_{m_s}+1}-t}{\alpha(m_s,j_s+1,d_s)}\mathfrak{B}_{j_s+1,d_{m_s}-1}^{(m_s)}(t),$$
where $\alpha(m_s,j_s,d_{m_s}) = \lambda^*_{m_s,j_s+d_{m_s}} - \lambda^*_{m_s,j_s}$. Using this formulation, 
\begin{equation}
\begin{array}{lcl}   
I(r, \bm, \bj, \bd) & = & \frac{1}{\alpha(m_{\bar{s}},j_{\bar{s}},d_{m_{\bar{s}}})}I(r+1,\bm,\bj,\bd-\be_{\bar{s}}) \\
& & -\frac{\lambda^*_{m_{\bar{s}},j_{\bar{s}}}}{\alpha(m_{\bar{s}},j_{\bar{s}},d_{m_{\bar{s}}})} I(r,\bm,\bj,\bd-\be_{\bar{s}}) \\
& & +\frac{\lambda^*_{m_{\bar{s}},j_{\bar{s}}+d_{m_{\bar{s}}}+1}}{\alpha(m_{\bar{s}},j_{\bar{s}}+1,d_{m_{\bar{s}}})} I(r,\bm,\bj,\bd-\be_{\bar{s}})\\
& & - \frac{1}{\alpha(m_{\bar{s}},j_{\bar{s}}+1,d_{m_{\bar{s}}})} I(r+1,\bm,\bj,\bd-\be_{\bar{s}}),
\end{array}
\label{eqn:recursion}
\end{equation}
where $\bar{s}$ is any $s=1,\dots,S$, such that $d_{m_{\bar{s}}} > 0$, and $\be_{\bar{s}}$ is the $S \times 1$ vector of zeros, apart from the $\bar{s}$th element, which is one. The recursion in (\ref{eqn:recursion}) continues until $I(r, \bm, \bj, \bd)$ can be written as a linear combination of integrals where the integrand features a product of zero-degree B-spline basis functions, i.e. integrals of the form 
\begin{eqnarray*}
I(r, \bm, \bj, \bzero) &=& \int_0^T t^r \prod_{s=1}^S \mathfrak{B}_{j_s,0}^{(m_s)}(t) \mathrm{d}t\\
 & = & \left\{
 \begin{array}{ll}
 \frac{\left[ L - U \right]^{r+1}}{r+1} & \mbox{if $L > U$;}\\
0 &  \mbox{if otherwise;} 
\end{array}\right.
\end{eqnarray*}
where $U = \max_{s=1,\dots,S} \left\{ \lambda^*_{m_s,j_s}\right\}$ and $L=\min_{s=1,\dots,S} \left\{ \lambda^*_{m_s,j_s+1} \right\}$.

\section{Proof of LEMMA 1}

The model matrix $Z$ can be written as $Z = (Z_1,\dots,Z_Q)$, where, for $q=1,\dots,Q$
$$Z_q = \left( \begin{array}{c}
\bz_{1q}^\T\\
\vdots \\
\bz_{nq}^\T \end{array}\right)
= \left( \begin{array}{c}
\bigotimes_{a \in \mathcal{F}_q} \bgamma_{1a}^\T \\
\vdots \\
\bigotimes_{a \in \mathcal{F}_q} \bgamma_{na}^\T \end{array} \right) R_q^\T
$$
is an $n \times m_{\beta,q}$ matrix. For the functional parameter coefficients, $\btheta$, to be estimable, $Z^\T Z$ is required to be non-singular, and it follows that $Z_q$ needs to have full column-rank, for all $q=1,\dots,Q$. Using properties of the rank of products of matrices \citep[e.g.][page 80]{gentle2007},
\begin{eqnarray}
\rank (Z_q)  & \le &  \min \left[ \rank(S), \rank(R_q^\T) \right] \nonumber \\
& \le & \min \left[ \min \left( n, \prod_{a \in \mathcal{F}_q} m_{x,a} \right), \min \left( \prod_{a \in \mathcal{F}_q} m_{x,a}, m_{\beta,q}\right) \right] \label{eqn:rank}
\end{eqnarray}
where 
$$S = \left( \begin{array}{c}
\bigotimes_{a \in \mathcal{F}_q} \bgamma_{1a}^\T \\
\vdots \\
\bigotimes_{a \in \mathcal{F}_q} \bgamma_{na}^\T \end{array} \right).$$
The right-hand-side of (\ref{eqn:rank}) is equal to $m_{\beta,q}$, which is required for $Z_q$ to be full column-rank, if and only if $m_{\beta,q} \le \prod_{a \in \mathcal{F}_q} m_{x,a}$.

\section{Additional details for the illustrative example in Section 4}

In this section, expressions are provided for the two matrices, $R_1$ and $R_2$, which feature in the model matrix $Z$, as well as the weighting matrix, $B_I$, for WNSEL Bayesian designs and the matrix, $G$, featuring in the roughness penalty.

The regression functional, $\bff(\cdot)$ means that $\mathcal{F}_1 = \varnothing$ and $\mathcal{F}_2 = \left\{1\right\}$. In both Cases I and II, $R_1$ is a scalar given by
$$R_1 = \int_0^1 b_1(t) \; \mathrm{d}t = 1.$$

\subsection{Matrices for Case I}

The $R_2$ matrix is $2 \times m_{x,1}$ and is given by
\begin{eqnarray*}
R_2 & = & \int_0^1 \left(\begin{array}{c}
1 \\
t \end{array} \right) \bc_1(t) \; \mathrm{d}t \\
& = & \left( \begin{array}{lllll}
\int_0^{\lambda_{1,1}} \; \mathrm{d}t & \int_{\lambda_{1,1}}^{\lambda_{1,2}} \; \mathrm{d}t & \dots & \int_{\lambda_{1,m_{x,1}-2}}^{\lambda_{1,m_{x,1}-1}} \; \mathrm{d}t & \int_{\lambda_{1,m_{x,1}-1}}^1 \; \mathrm{d}t
\\
\int_0^{\lambda_{1,1}} t\; \mathrm{d}t & \int_{\lambda_{1,1}}^{\lambda_{1,2}}t \; \mathrm{d}t & \dots & \int_{\lambda_{1,m_{x,1}-2}}^{\lambda_{1,m_{x,1}-1}}t \; \mathrm{d}t & \int_{\lambda_{1,m_{x,1}-1}}^1t \; \mathrm{d}t
\end{array}\right)\\
& = & 
\left( \begin{array}{lllll}
\lambda_1 & \lambda_2-\lambda_1 & \dots & \lambda_{1,m_{x,1}-1}-\lambda_{1,m_{x,1}-2} & 1-\lambda_{1,m_{x,1}-1}\\
\frac{1}{2}\lambda_1^2 & \frac{1}{2}(\lambda_2^2-\lambda_1^2) & \dots & \frac{1}{2}(\lambda_{1,m_{x,1}-1}^2-\lambda_{1,m_{x,1}-2}^2) & \frac{1}{2}(1-\lambda_{1,m_{x,1}-1}^2)
\end{array} \right).
\end{eqnarray*}
The $B_I$ matrix is $3 \times 3$ and is given by
\begin{eqnarray*}
B_I & = & \int_0^1 B(t) B(t)^\T \; \mathrm{d}t\\
& = & \left[ \begin{array}{ccc}
1 & 0 & 0 \\
0 & 1 & \frac{1}{2}\\
0 & \frac{1}{2} & \frac{1}{3} \end{array}\right],
\end{eqnarray*}
which follows from 
$$B(t) = \left(\begin{array}{cc}
b_1(t) & 0 \\
0 & \bbb_2(t)
\end{array}\right) = \left(\begin{array}{cc}
1 & 0 \\
0 & 1 \\
0 & t
\end{array}\right).
$$

\subsection{Matrices for Case II}

The $R_2$ matrix is $3 \times m_{x,1}$ and is given by
\begin{eqnarray*}
R_2 & = & \int_0^1 \left(\begin{array}{c}
1 \\
t \\
t^2
\end{array} \right) \bc_1(t) \; \mathrm{d}t \\
& = & \left( \begin{array}{lllll}
\int_0^{\lambda_{1,1}} \; \mathrm{d}t & \int_{\lambda_{1,1}}^{\lambda_{1,2}} \; \mathrm{d}t & \dots & \int_{\lambda_{1,m_{x,1}-2}}^{\lambda_{1,m_{x,1}-1}} \; \mathrm{d}t & \int_{\lambda_{1,m_{x,1}-1}}^1 \; \mathrm{d}t
\\
\int_0^{\lambda_{1,1}} t\; \mathrm{d}t & \int_{\lambda_{1,1}}^{\lambda_{1,2}}t \; \mathrm{d}t & \dots & \int_{\lambda_{1,m_{x,1}-2}}^{\lambda_{1,m_{x,1}-1}}t \; \mathrm{d}t & \int_{\lambda_{1,m_{x,1}-1}}^1t \; \mathrm{d}t \\
\int_0^{\lambda_{1,1}} t^2\; \mathrm{d}t & \int_{\lambda_{1,1}}^{\lambda_{1,2}}t^2 \; \mathrm{d}t & \dots & \int_{\lambda_{1,m_{x,1}-2}}^{\lambda_{1,m_{x,1}-1}}t^2 \; \mathrm{d}t & \int_{\lambda_{1,m_{x,1}-1}}^1t^2 \; \mathrm{d}t \end{array}\right)\\
& = & 
\left( \begin{array}{lllll}
\lambda_1 & \lambda_2-\lambda_1 & \dots & \lambda_{1,m_{x,1}-1}-\lambda_{1,m_{x,1}-2} & 1-\lambda_{1,m_{x,1}-1}\\
\frac{1}{2}\lambda_1^2 & \frac{1}{2}(\lambda_2^2-\lambda_1^2) & \dots & \frac{1}{2}(\lambda_{1,m_{x,1}-1}^2-\lambda_{1,m_{x,1}-2}^2) & \frac{1}{2}(1-\lambda_{1,m_{x,1}-1}^2)\\
\frac{1}{3}\lambda_1^3 & \frac{1}{3}(\lambda_2^3-\lambda_1^3) & \dots & \frac{1}{3}(\lambda_{1,m_{x,1}-1}^3-\lambda_{1,m_{x,1}-2}^3) & \frac{1}{3}(1-\lambda_{1,m_{x,1}-1}^3)
\end{array} \right).
\end{eqnarray*}

The $B_I$ matrix is $4 \times 4$ and is given by
\begin{eqnarray*}
B_I & = & \int_0^1 B(t) B(t)^\T \; \mathrm{d}t\\
& = & \left[ \begin{array}{cccc}
1 & 0 & 0 & 0\\
0 & 1 & \frac{1}{2} & \frac{1}{3}\\
0 & \frac{1}{2} & \frac{1}{3} & \frac{1}{4} \\
0 & \frac{1}{3} & \frac{1}{4} & \frac{1}{5} 
\end{array}\right],
\end{eqnarray*}
which follows from 
$$B(t) = \left(\begin{array}{cc}
b_1(t) & 0 \\
0 & \bbb_2(t)
\end{array}\right) = \left(\begin{array}{cc}
1 & 0 \\
0 & 1 \\
0 & t \\
0 & t^2
\end{array}\right).
$$




\section{Sensitivity analyses for Ambr250 bioreactor experiment} \label{sec:sensitivity}

Table~\ref{tab:sens} shows the objective function values for SE optimal designs for $m_{x,1}=3,4,8,16,100$.

\begin{table}[ht] 
\centering
\caption{SE objective function values for optimal designs for $m_{x,1}=3,4,8,16,100$.} \label{tab:sens}
\begin{tabular}{ cc }
 \hline
$m_{x,1}$ & $n=12$ \\
\hline
3    & 128.802    \\
4    &  69.802    \\
8    &  68.085    \\
16   &  66.505    \\
100  &  65.304    \\
\hline
\end{tabular}
\end{table}

\section{Choice of the prior through a roughness penalty} \label{sec:roughness}

The choice of basis functions is typically challenging. For instance, for a B-spline of fixed degree, choosing the number and placement of the knots is equivalent to the number of basis functions. An alternative approach is to use a roughness penalty, which enforces smoothing by penalising the complexity of the functional parameters. The roughness penalty helps prevent overfitting and simplifies the problem to choosing a single scalar smoothing parameter which controls the trade-off between fit and smoothness. 

A roughness penalty can be implemented through the prior variance matrix as $\Sigma^{-1} = \alpha G$, where $\alpha > 0$ is the smoothing parameter and $G$ is a $p \times p$ matrix. Larger values of $\alpha$ lead to higher roughness penalties (smoother functional parameters). 

Let $B''(t)$ be the $p \times Q$ matrix with $iq$th element $B''_{iq}(t) = \partial^2 B_{iq}(t)/\partial t^2$, for $q=1,\dots,p$ and $q=1,\dots,Q$. Then, following, for example \cite[][Chapter 5]{ramsay2005}, the $ij$th element of $G$ is $G_{ij} = \int_{\mathcal{T}} U_{ij}(t) \; \mathrm{d}t$ where $U_{ij}$ is the $ij$th element of the $p \times p$ matrix given by $U = B''(t)B''(t)^\T$. 

If $\alpha = 0$ or the functional parameters are constants, then there is no roughness penalty, and the SE and WSE Bayesian optimal designs are equivalent to classical A- and L-optimal designs, respectively.

\end{document}